\documentstyle[prl,aps,floats,psfig]{revtex}

\begin{document}
\widetext
\draft
\twocolumn[\hsize\textwidth\columnwidth\hsize\csname @twocolumnfalse\endcsname
\title{Direct Numerical Observation of Anomalous Diffusion in Random Media
}

\author{
Victor Pham
and
Michael W. Deem
}

\address{
Chemical Engineering Department, University of
California, Los Angeles, CA  90095-1592\\
mwdeem@ucla.edu
}
\maketitle

\begin{abstract}
We present computer simulations of anomalous diffusion, 
$\langle r^2(t) \rangle \sim a t^{1-\delta}$, in two
dimensions.  The Monte Carlo calculations are in excellent
agreement with previous renormalization group calculations.
Interestingly,  use of a high quality pseudo-random number
generator is necessary to observe the anomalous diffusion.
A linear-feedback, shift-register method
leads to incorrect, super-diffusive motion of the random walkers.
\end{abstract}
\pacs{82.20.Db, 05.40.+j, 82.20.Mj}
]

\narrowtext
\section*{1. Introduction}
Certain types of physically realizable disorders cause anomalous sub-diffusion
in two dimensions
\cite{Fisher,Kravtsov1}.
In these unusual cases the mean square displacement of random walkers
does not increase linearly with time, but rather increases sub-linearly
with time, $\langle r^2(t) \rangle \sim a t^{1-\delta}$.  The
exponent for this scaling is continuously variable in the strength
of disorder.  In fact, the exponent can be found exactly for the type of
singular Gaussian disorder that leads to sub-diffusion
\cite{Kravtsov2,Bouchaud1,Honkonen1,Honkonen2,Derkachov1,Derkachov2}.
  This special disorder corresponds physically to quenched,
charged defects.  The defects can either be true, electrostatic charges or
topological ``charges.''  In either case, the quenched charges interact
with the moving particles of interest via a long-range, logarithmic
potential.

Diffusion in a variety of random media has been considered by
numerical simulation.
Typical disorders that have been investigated include
non-singular potential disorder \cite{Pautmeier}, random fluid
flows \cite{Mosler,Dean,Elliott},
 fractal media
\cite{Elliott2,IV_Havlin,IV_Isichenko}, optical molasses \cite{Marksteiner},
and topologically disordered structures \cite{Bausch}.
The random walk Monte Carlo simulation method is used in 
most numerical studies of diffusion.  
To date, however, there are no numerical
studies that verify the renormalization group predictions  of anomalous 
diffusion in singular, two-dimensional, random potential fields.
 Preliminary results were cited in the review by Bouchaud and Georges
\cite{Bouchaud3}, but a formal publication did not ensue.

 This same type of quenched, ``ionic'' disorder
subjects chemical reactions to 
transport limitations and causes anomalous kinetic behavior.  
  Two dimensions is particularly interesting, since
this is also the upper critical dimension for bimolecular reactions
\cite{Peliti,Lee2,Deem1,Deem2}, in addition to being the upper critical
dimension for the charged disorder.
The kinetics of chemical reactions, of course, have always been a subject of
interest for scientists and engineers.
In most cases, the reactant diffusion is normal,
and the effect of diffusion limitations
on the rate of chemical reaction is easy to calculate.
Anomalous diffusion, however, leads to anomalous kinetics.
In this case, the effect on the rate of chemical reaction is
not so widely known.
The kinetics of the reactions 
$A+A \to \emptyset$,
$A+B \to \emptyset$, and
$A^+ + B^-
~{\mathrel{\mathop{\rightleftharpoons}\limits^{\lambda}_{\tau}}}~
AB$
in singular disorder
have been derived analytically \cite{Deem1,Deem2,Deem3}.
In these references, 
 a field-theoretic treatment of
anomalous kinetics was worked out, and renormalization group
predictions were derived.  These studies show that the reactions
become transport limited in the long time regime.  At long times,
where the diffusion is anomalous, 
the kinetics also becomes anomalous.

Simulations to test these theoretical predictions would be of
great interest.
A necessary prerequisite for proceeding
with these numerical studies of anomalous kinetics is first the
ability to
simulate anomalous diffusion.  This ability
requires both constructing the disorder and
performing the diffusive motion in the quenched disorder.

In this article, we
present numerical observations of anomalous diffusion in two
dimensions using the Monte Carlo method. 
 In section 2 we discuss the appropriate form of the
quenched disorder.
In  section 3 we
introduce our method for creating the random potential
and for simulating motion in this potential.
 Our results are
presented in section 4.
A discussion of the results is presented in
section 5, where a comparison is made
with the renormalization group predictions. 
We conclude in section 6.

\section*{2. The Quenched Disorder}
We consider the motion of one charged particle in a sea
of quenched charges in two dimensions.  The statistics of the
motion of the particle is completely determined by the statistics of
the quenched potential field that the particle encounters.
The quenched charges, which obey bulk charge neutrality, give rise
to a charge-charge correlation function that vanishes as $k^2$ for
small $k$ in Fourier space.  The potential, which is the convolution of
the charge distribution with the logarithmic Coulomb law, gives rise
to a potential-potential correlation function that diverges as
$1/k^2$ for small $k$ \cite{Deem0}.  This long-ranged correlation 
function for the potential experienced by the diffusing particle is
exactly of the form that leads to anomalous diffusion.  

In two dimensions the interaction between the charges and the diffusing
particle, and between the quenched charges themselves, is logarithmic.
Electrostatic charges 
in two dimensions, or line charges in three dimensions, interact
with this law.  Certain topological defects in two dimensions also interact
with this same law.  For example, dislocations in solids 
interact with a logarithmic law due to induced long-range
elastic strain fields.  Disclinations in hexatic membranes also
interact with this effective law, due to screening of the
induced strain fields by free dislocations.
Typical examples of these topological defects include line defects
in three dimensional crystals,
vortices in superfluids, and
flux lines in superconductors \cite{NelsonII}.

The exponent for the mean square displacement depends indirectly on the
density of defects via the prefactor of the potential-potential
correlation function.  The form of correlation function appropriate for
small $k$ is
$\hat{\chi}_{vv}(k) \sim \gamma/k^{2}$, 
where $\gamma$ is the strength of the disorder, and
$\hat \chi_{vv}({\bf k}) = \int d^d {\bf r} \exp(i {\bf k}\cdot{\bf r})
\chi_{vv}({\bf r})$ is the Fourier transform of the correlation
function.
We are free
to chose different behavior away from the origin in Fourier space.
A natural choice for this correlation function
 is the inverse of the diffusive Green's function.
On a square lattice with spacing $\Delta r$ we, thus, use the form
\begin{equation}
\hat{\chi}_{vv}({\bf k})=\frac{\gamma (\Delta r)^{2}}{4-2\cos(k_{x} \Delta
r)-2\cos(k_{y} \Delta r)} \ .
\label{1}
\end{equation}
Note that this form of $\hat{\chi}_{vv}(k)$ has the appropriate
limiting behavior as the lattice spacing
become infinitesimally small and as the wavelength becomes large.
The particles move through this potential field  starting from random initial
positions. Renormalization group calculations have rather convincingly shown
that the mean square
displacement exhibits an anomalous behavior at long times
\cite{Fisher,Kravtsov1,Kravtsov2,Bouchaud1,Honkonen1,Honkonen2,Derkachov1,Derkachov2,Bouchaud2}.
  On a log-log scale, the mean square
displacement as a function of time has a slope of $1-\delta$, where
\begin{equation}
\delta = \left[ 1+\frac{8\pi}{\beta^{2}\gamma} \right]^{-1}
\label{2}
\end{equation}
and $\beta=1/(k_{B}T)$.

\section*{3. Simulation Method}
We now consider the creation of the Gaussian random
potential $V({\bf r})$
on a lattice.
The potential takes on real values at each lattice site.  The
probability of observing any specific potential distribution is
given by
\begin{equation}
P[V] = \frac{e^{-\beta H [V]}}{ Z} \ ,
\label{3}
\end{equation}
where
\begin{eqnarray}
\beta H [V]& = & \frac{1}{2} \int d^{d}{\bf r} d^{d}{\bf r}' V({\bf r}) 
             \chi_{vv}^{-1}( {\bf r} - {\bf r}' ) V({\bf r}')  \nonumber \\
        & = & \frac{1}{2} \int \frac{d^{d}\bf k}{(2\pi)^d}|\hat{V}({\bf k})|^{2} 
             \hat{\chi}_{vv}^{-1}({\bf k})  \nonumber \\
        & = & \frac{1}{2} \sum_{\bf k} \frac{(\Delta k)^{d}}{(2\pi)^{d}}
             |\hat{V}({\bf k})|^{2} \hat{\chi}_{vv}^{-1}({\bf k}) \ ,
\label{4}
\end{eqnarray}
$Z$ is a normalizing constant, and $\Delta k=2 \pi/(N \Delta r)$.  
Here the lattice is in $d$ dimensions ($d=2$ in our case), has $N$
unit cells on a side, and has lattice spacing $\Delta r$.  The
Fourier transform is given by $\hat V({\bf k}) = \int d^d {\bf r}
V({\bf r}) \exp(i {\bf k} \cdot {\bf r}) = \sum_{\bf r} (\Delta r)^d
V({\bf r}) \exp(i {\bf k} \cdot {\bf r})$.
 Since $\hat{V}(-{\bf k})=\hat{V}^{*}({\bf k})$,  we have
\begin{equation}
\beta H = \sum_{{\bf k}_{1/2}} \sigma({\bf k}) \left( \frac{\Delta k}{2\pi}
\right)^{d} |\hat{V}({\bf k})|^{2} \hat{\chi}_{vv}^{-1}({\bf k})
\label{5}
\end{equation}
with ${\bf k}_{1/2}$ meaning half of $k$ space and
\begin{equation}
\sigma({\bf k}) =  \left\{
                   \begin{array} {ll}
                      1/2,& \mbox{if ${\bf k}$ is on a special point} \\[.2in]
                      1,  & \mbox{otherwise}
                   \end{array}
                   \right. \ ,
\label{6}
\end{equation}
where the special points are the origin, the corners of the lattice,
and the intersections of $k_{x}$-axis and $k_{y}$-axis with the lattice
boundaries.  In this form, $\hat{V}({\bf k})$ and $\hat{V}({\bf k}')$
are independent as long as ${\bf k} \neq {\bf k}'$. The potential
$\hat{V}({\bf k})$
is Gaussian with the following variance for the real and imaginary
components for ${\bf k}$ not special
\begin{equation}
\left\langle \left[ {\rm Re}\hat{V}({\bf k}) \right]^{2} \right\rangle = 
\frac{1}{2}\Omega \hat{\chi}_{vv}({\bf k})
\label{7}
\end{equation}
\begin{equation}
\left\langle \left[ {\rm Im}\hat{V}({\bf k}) \right]^{2} \right\rangle = 
\frac{1}{2} \Omega \hat{\chi}_{vv}({\bf k}) \ ,
\label{8}
\end{equation}
where $\Omega = (N \Delta r)^d$ is the volume of the system.
If ${\bf k}$ is special,
\begin{equation}
{\rm Im}\hat{V}({\bf k}) = 0 \ ,
\label{9}
\end{equation}
and 
\begin{equation}
\left\langle \left[ {\rm Re}\hat{V}({\bf k}) \right]^{2} \right\rangle = 
\Omega \hat{\chi}_{vv}({\bf k}) \ .
\label{10}
\end{equation}
Also, by definition, $\hat{V}({\bf 0}) = \int {d}^{d}{\bf r} V({\bf r})$.
Since it is not the actual magnitude of the potential but rather its
gradient that is of interest, we define $V' = V - \langle V \rangle$.
In this form, $\hat{V}'({\bf 0}) = 0$ and $\nabla V' = \nabla V$.
We use the potential $V'$ in the simulation.  To create the potential
$V'({\bf r})$ that we need, 
we first create $\hat V'({\bf k})$ in half of k-space by generating independent
Gaussian random numbers with variances
given by
Eqs.\ (\ref{7})-(\ref{10}).  We then generate the other half of
k-space using the relation $\hat V'(-{\bf k}) = \hat V'^*({\bf k})$.
Finally, we generate $V'$ in real space by performing an inverse
fast Fourier transform \cite{Press}.

Alternatively, we can construct a  real potential by first 
generating a complex $V({\bf r})$ and then extracting a real potential,
$V'({\bf r})$.
Specifically, we generate a complex Gaussian field
with the probability distribution
\begin{equation}
P[V] = Z^{-1} \exp\left[-\int d^{d}{\bf r}  d^{d}{\bf r'} V^{*}({\bf r}) \chi^{-1}({\bf r}
-{\bf r}') V({\bf r}')\right] \ .
\label{11}
\end{equation}
Note that the fields
${\rm Re} V({\bf r})$ and ${\rm Im} V({\bf r})$ are each Gaussian. Thus,
we find
\begin{equation}
\langle V^{*}({\bf r})V({\bf r}') \rangle = \chi({\bf r}-{\bf r}').
\label{12}
\end{equation}
We define
\begin{equation}
V'({\bf r}) = {\rm Re} V({\bf r}) + {\rm Im} V({\bf r}).
\label{13}
\end{equation}
$V'({\bf r})$ is also Gaussian, since
\begin{eqnarray}
  & \int d^{d}{\bf r} d^{d}{\bf r}' V^{*}({\bf r}) \chi_{vv}^{-1}({\bf r}-{\bf r}') 
    V({\bf r}')  \nonumber \\
= & \int d^{d}{\bf r} d^{d}{\bf r}' {\rm Re}V({\bf r}) 
    \chi_{vv}^{-1}({\bf r}-{\bf r}') {\rm Re}V({\bf r}')  \nonumber \\
+ & \int d^{d}{\bf r} d^{d}{\bf r}' {\rm Im}V({\bf r}) 
    \chi_{vv}^{-1}({\bf r}-{\bf r}') {\rm Im}V({\bf r}') \nonumber  \\
- & i \int d^{d}{\bf r} d^{d}{\bf r}' {\rm Im}V({\bf r}) 
      \chi_{vv}^{-1}({\bf r}-{\bf r}') {\rm Re}V({\bf r}') \nonumber \\ 
+ & i \int d^{d}{\bf r} d^{d}{\bf r}' {\rm Re}V({\bf r}) 
      \chi_{vv}^{-1}({\bf r}-{\bf r}') {\rm Im}V({\bf r}').
\label{14}
\end{eqnarray}
The complex part of the above equation vanishes, since
$\chi_{vv}^{-1}({\bf r}-{\bf r}')=\chi_{vv}^{-1}({\bf r}'-{\bf r})$
for a medium obeying Eq.\ (\ref{1}).
From our definition of $V'$, we find
\begin{equation}
\langle V'({\bf r})V'({\bf r}') \rangle = \chi_{vv}({\bf r}-{\bf r}').
\label{15}
\end{equation}
This definition of $V'$ leads to a real, Gaussian potential field with
the correct statistics.  It is, therefore, an equally valid
quenched random potential.

With the potential established, we shift focus to the
initial conditions for the particle and the
mechanics for diffusion. In the theoretical
treatment of anomalous diffusion, the particles are assumed to be
uniformly distributed over the lattice. In our simulation, therefore,
we choose the initial positions of the random walkers from a uniform
distribution. In addition, we also examine the case of a Boltzmann
distribution of initial positions, as would be appropriate, for
example, for a NMR
experiment on an equilibrated system.  
The particles are
selected from a random initial position on a finite lattice, and they
begin to
diffuse at $t=0$.  We could let one particle diffuse an infinitely long
time and record its behavior exactly as prescribed by theory.
Depending on whether the diffusion is self-averaging, we may also need
to average over different realizations of the potential.
Unfortunately,
lattice effects will appear at long
times as a result of the periodic
boundary conditions. Another method, which is more efficient and 
yields better
statistics, is to sample many random walkers for a shorter period
of time.
 This will work as long as the observation time is long enough to be in the
scaling region, that is, as long as
the system is large enough. We have investigated
finite size effects and found that a square lattice of 
$N=2048$ unit cells on a side
is sufficiently large to allow the particles to be in the
scaling regime for our range of parameter values.
At short times, the particles will display normal diffusion behavior.
At intermediate times, the diffusivity will tend to zero.
At long times, we will observe the anomalous sub-diffusion,
where $\langle r^{2}(t) \rangle$ is
expected to be proportional to $t^{1-\delta}$ as $t \rightarrow
\infty$.

The statistics of the random walk on the lattice are conveniently described
by a master equation.  This master equation defines the probabilities and
rates of all possible hops that the random walker can execute.
We exactly solve this master  equation by a Poisson process \cite{Tau}.
In one dimension, the master equation looks like
\begin{equation}
\frac{dP_{n}}{dt} = U_{n-1}P_{n-1}-D_{n}P_{n}+D_{n+1}P_{n+1}-U_{n}P_{n},
\label{16}
\end{equation}
where $P_{n}$ is the probability for being at site $x_{n}$ at time
$t$, $U_{n}$ is the rate at which transitions occur from $x_{n}$ to
$x_{n+1}$ and $D_{n}$ is the rate at which transitions occur from
$x_{n}$ to $x_{n-1}$.  In a two dimensional space, we denote the rates
by $\tau_{n,m}^{(1)}$, $\tau_{n,m}^{(2)}$, $\tau_{n,m}^{(3)}$, and
$\tau_{n,m}^{(4)}$, where $\tau_{n,m}^{(1)}$ and $\tau_{n,m}^{(2)}$
are rates at which transitions occur from $x_{n,m}$ to $x_{n+1,m}$ and
$x_{n-1,m}$ respectively, and $\tau_{n,m}^{(3)}$ and
$\tau_{n,m}^{(4)}$ 
are rates at which transitions occur 
from $x_{n,m}$ to $x_{n,m+1}$ and $x_{n,m-1}$
respectively. The two dimensional master equation  is
\begin{eqnarray}
\frac{dP_{n,m}}{dt} & = & \tau_{n-1,m}^{(1)}P_{n-1,m}-\tau_{n,m}^{(1)}P_{n,m} \nonumber  \\
& + & \tau_{n+1,m}^{(2)}P_{n+1,m} -\tau_{n,m}^{(2)}P_{n,m}  \nonumber  \\
   & + & \tau_{n,m-1}^{(3)}P_{n,m-1}-\tau_{n,m}^{(3)}P_{n,m} \nonumber  \\
 &+ &\tau_{n,m+1}^{(4)}P_{n,m+1}-\tau_{n,m}^{(4)}P_{n,m}.
\label{17}
\end{eqnarray}
We demand that this master equation lead to a Boltzmann distribution
of the random walkers at long times.  In other words, we want the
stationary probability $P_{n}^{S}$ of being at site $x_{n,m}$ to be
\begin{equation}
P_{n}^{S} = \frac{(\Delta r)^{2} e^{-\beta V(x_{n,m})}}{Z}.
\label{18}
\end{equation}
This distribution will arise if detailed
balance is enforced.  For example, the condition of detailed balance
for transitions in the x direction  is
\begin{equation}
\tau_{n+1,m}^{(2)}P_{n+1,m} = \tau_{n,m}^{(1)}{P_{n,m}} \ .
\label{19}
\end{equation}
This criterion implies, with the use of the Boltzmann distribution,
\begin{equation}
\frac{P_{n+1,m}}{P_{n,m}} = \frac{\tau_{n,m}^{(1)}}{\tau_{n+1,m}^{(2)}} = 
e^{-[\beta V(x_{n+1,m})-V(x_{n,m})]} \ .
\label{20}
\end{equation}
Hence, one 
consistent expression for the 
rates $\tau_{n,m}^{(1)}$ and $\tau_{n,m}^{(2)}$ is
\begin{eqnarray}
\tau_{n,m}^{(1)} & = & \frac{D}{ (\Delta r)^{2}}e^{\beta [V(x_{n,m})-V(x_{n+1,m})]/2} \\
\tau_{n,m}^{(2)} & = & \frac{D}{(\Delta r)^{2}}e^{\beta [V(x_{n,m})-V(x_{n-1,m})]/2} \ .
\label{21}
\end{eqnarray}
Similar relations hold for the transition rates
in the y direction:
\begin{eqnarray}
\tau_{n,m}^{(3)} & = & \frac{D}{ (\Delta r)^{2}}e^{\beta [V(x_{n,m})-V(x_{n,m+1})]/2} \\
\tau_{n,m}^{(4)} & = & \frac{D}{(\Delta r)^{2}}e^{\beta [V(x_{n,m})-V(x_{n,m-1})]/2} \ .
\label{21a}
\end{eqnarray}

  We use these transition rates to generate
a stochastic Poisson
process. Specifically, a particle begins at a site
$x_{n,m}$. The particle waits at this site with an exponentially
distributed amount of time characterized by its mean value
\begin{equation}
\langle d t  \rangle = \frac{1}{\tau_{n,m}^{(1)}+ \tau_{n,m}^{(2)} + \tau_{n,m}^{(3)} + \tau_{n,m}^{(4)}} \ .
\label{22}
\end{equation}
We generate the actual time increment via
\begin{equation}
d t=-\frac{1}{\tau_{n,m}^{(1)}+ \tau_{n,m}^{(2)} + \tau_{n,m}^{(3)} + \tau_{n,m}^{(4)}} \ln (x) \ ,
\label{23}
\end{equation}
where $x$ is uniformly distributed random number with $0 < x \le 1$.
We then generate a
second uniformly 
distributed random number that we use
to pick one of the four possible nearest-neighbor hops according to
their probabilities:
\begin{eqnarray}
P(x_{n,m} \rightarrow x_{n+1,m}) & = & 
\frac{\tau_{n,m}^{(1)}}
{\tau_{n,m}^{(1)} +
 \tau_{n,m}^{(2)} + 
 \tau_{n,m}^{(3)} + 
 \tau_{n,m}^{(4)} } \nonumber \\
P(x_{n,m} \rightarrow x_{n-1,m}) & = & 
  \frac{\tau_{n,m}^{(2)}}
  {\tau_{n,m}^{(1)}+ 
  \tau_{n,m}^{(2)} + 
  \tau_{n,m}^{(3)} + 
  \tau_{n,m}^{(4)}}\nonumber \\
P(x_{n,m} \rightarrow x_{n,m+1}) & = & 
\frac{\tau_{n,m}^{(3)}}
 {\tau_{n,m}^{(1)}+ 
  \tau_{n,m}^{(2)} + 
  \tau_{n,m}^{(3)} + 
  \tau_{n,m}^{(4)}}\nonumber \\
P(x_{n,m} \rightarrow x_{n,m-1}) & = & 
\frac{\tau_{n,m}^{(4)}}
{\tau_{n,m}^{(1)}+ 
\tau_{n,m}^{(2)} + 
\tau_{n,m}^{(3)} + 
\tau_{n,m}^{(4)}} \ .
\label{24}
\end{eqnarray}
The particle hops to the new site, and time is incremented by $d t$.

On a finite lattice with periodic boundary conditions, 
particles reenter the lattice when they reach the boundaries.
Furthermore, the shortest path between two particles may
cross the boundary.  This is significant, because we
are fundamentally interested in the implications of anomalous
diffusion for chemical reaction, and the appropriate measure
for defining distance between two reactants is the length of the shortest
path.
We, therefore, define the
distance traveled by a diffusing particle as
\begin{equation}
r^{2} = \min_{p,q} \left\{\left[ i-i_0 + p N
	\right]^2 + \left[ j-j_0 +q N  \right]^2 \right\} (\Delta r)^2 \ .
\label{25}
\end{equation}
Here 
$i_{0}$ and $j_{0}$ are the initial position coordinates of the
particle, $N$ is the dimension of the lattice, and the minimum over
integer $p$ and $q$ mathematically defines the shortest path.

An important component of the simulation is the random number
generator. A desirable generator ensures that the correct
random statistics are 
used in lattice creation,  choice of initial 
particle positions, choice of hoping
directions, and generation of time increments.
In this study, we used two
random number generators: one that is a
 sum of three linear congruential generators 
\cite{Byte} and an
exclusive-or, linear-feedback shift-register method \cite{Knuth}.
Unless specified otherwise, all the data that are 
shown were generated using the
sum of three linear congruential generator method.

\section*{4. Results}
We are interested in the long
 time diffusive behavior of the particles diffusing on the disordered
lattice.
Since the diffusion is anomalous, the slope 
of $\ln \langle r^{2} \rangle$ {\em versus} $\ln t
$ will not be unity.  This slope will approach 0 as the
strength of disorder goes to $\infty$ and will approach 1
as the strength of disorder goes to 0.  We, therefore, examine
disorder strengths of varying magnitudes. We find that values
of $\beta^2 \gamma$ in the range of 1 to 20 produce convincing and
high quality results. For $\beta^2 \gamma$ too small, the slope
will be $1-\varepsilon$, with $\varepsilon$ smaller than the noise in
our simulation, making the anomalous diffusion difficult to
observe. When $\beta^2 \gamma$ is too large, strong fluctuations in the
particle behavior appear at short times 
because of significant lattice effects.  Within the
chosen range of $\beta^2 \gamma$, we observe well behaved curves that exhibit
distinct disorder strength dependent slopes.

We perform several simulations with different
initial starting positions for the random walker.
We collect these data as histograms of $\langle r^2(t) \rangle$ {\em versus}
$t$, with a temporal bin width of $t_0 = 1$.
When plotted on a logarithmic scale, 
these histograms will have  much more data for large $\ln (t/t_0)$
than for small $\ln (t/t_0)$.
In order to counteract this effect, 
we use an exponential sequence for
selecting data from the histogram when performing the fit
to determine the slope. 
Data for both short times, when the diffusion is not yet anomalous,
 and long times, when finite size effects are significant, was not
used in the fitting procedure.
This procedure leads to
widely spaced, independent data points.  A convenient byproduct of this
procedure is that
the standard error of the fit gives a reliable estimate of the 
error in the measured $\delta$.
These error bars are included in all of the figures.
Note that there could also be a systematic error in each simulation related to
the fact that only a single realization of the disorder is employed.
This systematic error seems to be small, as simulations with different
realizations of the disorder lead to values of $\delta$ that differ
approximately by the standard error of the fit.

We pick enough different starting positions and follow each simulation
long enough to ensure adequate statistics.
We found that $N_{do} = 10000$ particles is sufficient to produce
fairly smooth histograms for the mean square displacement.
We also found that $N_{len} = 2000000$ steps in each random walk is
enough for the particles to reach the asymptotic scaling regime.
We found that a lattice of size $N = 2048$ is sufficient to give 
results that span a broad range of times in the scaling regime.
In all simulations, the lattice spacing $\Delta r$ is set equal to unity,
which we can enforce by a spatial rescaling.
We also set the bare diffusivity equal to unity, which we
can similarly enforce by a temporal rescaling.  These rescalings will not affect
the scaling behavior at long times.  The
value of $\delta$ observed at long times, for example, is independent of
these bare values.

Figure 1 shows the slopes determined from simulations with
strengths of disorder $1 \le \beta^2 \gamma \le 20$.
\begin{figure}[t]
\centering
\leavevmode
\psfig{file=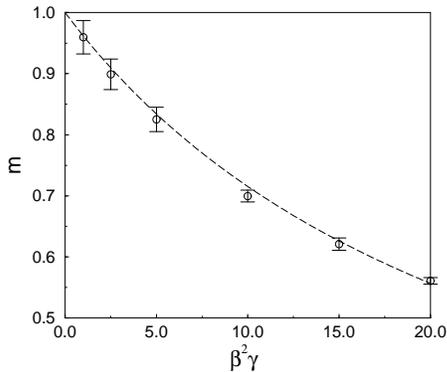,height=2in,angle=-90}
\caption[]
{\label{fig1}
Shown are the Monte Carlo results for the slope of the mean square
displacement, $m$, as a function of strength of disorder, $\beta^2 \gamma$.
 We have used Eq.
(\ref{5}) to generate the lattice and we placed the random walker
uniformly and randomly at the beginning of each random walk.
Shown in dashed are the renormalization group predictions.
}
\end{figure}
We have used Eq.\ (\ref{5}) to create the lattice in these simulations.
We have chosen the initial position of the random walker uniformly on
the lattice, in direct correspondence to the case considered by
the renormalization group studies.
The data points represent slopes of the
mean square travel distances of the random walkers as a function
of time.  We calculated these values by fitting
$\log \langle r^2(t)/(\Delta r)^2 \rangle$ as a function of $\ln (t/t_0)$,
as described above.
For each value of the disorder strength,  three independent
runs were performed on different realizations of the disorder.
We show in Figure 1
 the average slope (circles) and the associated standard error
of the fit (error bars) for each strength of disorder.
The renormalization group predictions are shown as the dashed line.
We see excellent agreement between the
simulation results and the theoretical predictions,
with nearly all the observed
values within one standard deviation of  the expected value.
The varying standard deviations reflect randomness in the potential fields,
initial positions, and hopping rates. The slopes do not include 
contributions from short time behavior or  long distance behavior 
[$\langle r^{2} \rangle$
on the order of $(N \Delta r)^2$].  Both of these regimes exhibit
 significant lattice effects, effects not considered in the renormalization
group studies.

Figure 2 shows similar data derived from lattice potentials constructed
from complex fields, Eq.\ (\ref{13}).  
\begin{figure}[t]
\centering
\leavevmode
\psfig{file=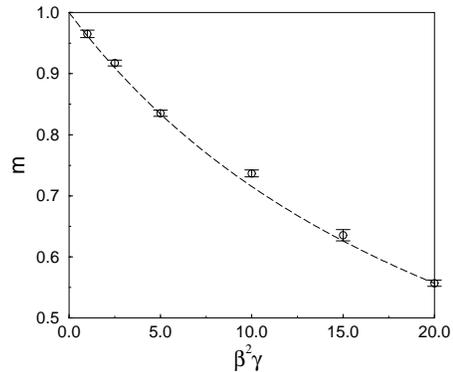,height=2in,angle=-90}
\caption[]
{\label{fig2}
The same quantities as in Figure 1, but using Eq.\
(\ref{13}) to generate the random potential lattice.
}
\end{figure}
These results should be identical to those of Figure 1.
We observe that the average values are, again, consistent with the
theoretical predictions.  Interestingly, the
standard errors are consistently
smaller than those of Figure 1.

Figure 3 shows the slopes that result when the particle initial positions
are chosen from a Boltzmann distribution.
\begin{figure}[t]
\centering
\leavevmode
\psfig{file=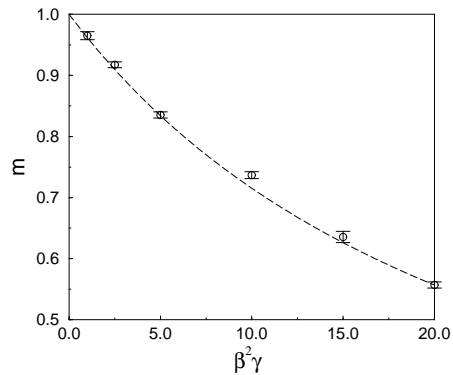,height=2in,angle=-90}
\caption[]
{\label{fig3}
The same quantities as in Figure 1, except that
we have placed the random walker randomly
according to a Boltzmann distribution at the beginning of each random walk.
}
\end{figure}
These conditions mimic those of a transient experiment,
such as pulsed field gradient NMR, performed on
an equilibrated system.
We see agreement between the observed values and the
renormalization group predictions, indicating that this 
change in initial conditions is not ``relevant'' in the technical sense.
We do see less
consistency in the standard deviations due to the
non-uniform sampling of the rugged potential landscape.

Figure 4 shows the slopes that result when one uses the correlation function
 \begin{equation}
\hat \chi_{vv}({\bf k}) = \gamma \frac{e^{-k^{2}/2}}{k^{2}} \ .
\label{26}
\end{equation}
\begin{figure}[t]
\centering
\leavevmode
\psfig{file=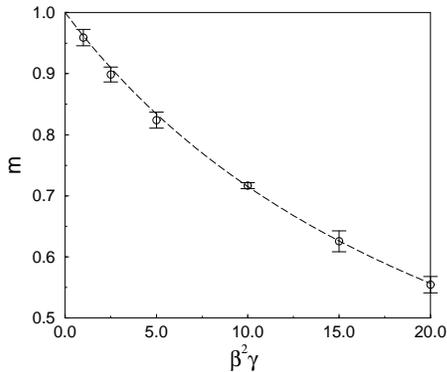,height=2in,angle=-90}
\caption[]
{\label{fig4}
The same quantities as in Figure 1, but using the
the correlation function $\hat \chi_{vv}(k) = \gamma \exp(-k^2/2)/k^2$
in creating the random potential.
}
\end{figure}
This correlation function has the same small $k$ behavior as 
Eq.\ (\ref{1}), but distinct behavior for large $k$.
The long-time behavior of these two correlation functions are
expected to be the same, as long as $\beta^2 \gamma$ is not
renormalized by changes in the large $k$ behavior.  
The prefactor  of the mean square displacement, however, is
observed to be
 significantly larger for Eq.\ (\ref{26}) than for Eq.\ (\ref{1}).

\section*{5. Discussion}
\begin{figure}[t]
\centering
\leavevmode
\psfig{file=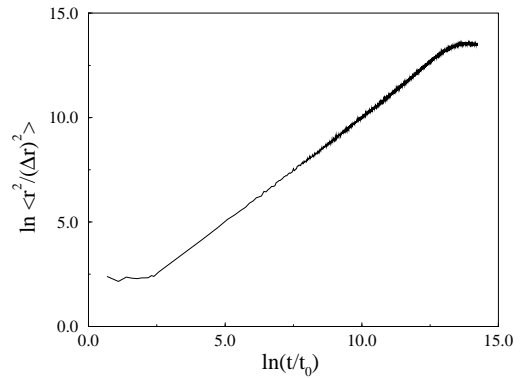,height=2in,angle=-90}
\caption[]
{\label{fig6}
Shown is the mean square displacement versus time for
$\beta^2 \gamma = 1$ and $\hat \chi_{vv}(k) =  \gamma  \exp(-k^2/2)/k^2$.
 We have used Eq.
(\ref{5}) to generate the lattice and we placed the random walker
uniformly and randomly at the beginning of each random walk.
Note the plateau in the mean square displacement at long times.
}
\end{figure}

The results of the computer simulations agree well with the
predictions of the
 renormalization group studies.
For each value of $\beta^2 \gamma $, the simulations yielded a
slope of $\ln \langle r^{2}/(\Delta r)^2 \rangle$ {\em versus} 
 $\ln (t/t_0) $ in excellent agreement with the analytical
predictions.  An important observation is that even at long times,
$\beta^2 \gamma $ does not become renormalized.
This is a non-trivial observation, as details in the simulation that
are technically ``irrelevant'' could renormalize $\beta^2 \gamma$ a 
finite amount.  For all values of $\beta^2 \gamma $,
lattice effects produce normal diffusive
behavior at short times.   This normal diffusion crosses over to
anomalous diffusion fairly quickly, within a time corresponding
to relatively few hops by the particle.
At very long times, the mean square displacement reaches a maximum
value due to the periodic boundary conditions.
This maximum
value is proportional to $N^2$, since the mean square displacement
defined in Eq.\ (\ref{25}) is always less than or equal to
$(N \Delta r)^2/2$.
    Figure
\ref{fig6} shows the mean square displacement measured in a typical
run.
The plateau at long times is clearly visible.  The normal diffusion
at short times crosses over to anomalous diffusion so rapidly that
it is not visible with the histogram bin width we used ($t_0 = 1$).

As discussed, both methods for generating the random lattice
potentials, Eq.\ (\ref{5}) and (\ref{13}),
lead to Gaussian random fields with the correct correlation function.
Both methods
should produce the same results for the mean square displacement.
We find that, indeed, both approaches give the same average result for
the mean square exponent.
The real lattice generation is
more efficient in terms of memory utilization.
One implication of this efficiency, however, 
is that the real lattice generation is constructed 
from fewer independent random numbers and is a more severe test of the
pseudo-random number generator.    The fewer degrees of freedom used
when implementing Eq.\ (\ref{3}) when compared to Eq.\ (\ref{13}) is
most likely what leads to the larger error bars 
in Figure 1 when compared to Figure 2.

The long-range correlations in the random potential allow,
in principle, for the
possibility that the mean-square displacement depends on the distribution
of initial conditions.  All of the theoretical predictions, for example,
are based upon the assumption of a uniform distribution of initial
conditions.  In experiments upon equilibrated systems, however, the
initial conditions are distributed in a Boltzmann manner.  We see
from Figure 3 that Boltzmann initial conditions lead to the same
mean square exponent at long times.

Two forms of the potential-potential correlation function, Eq.\ (\ref{1})
 and (\ref{26}), are used to explore the dependence on irrelevant,
large $k$ details.
At large distances and long times, the observed scaling
behavior that results from the two
correlation functions  would differ only if technically irrelevant details
of the lattice renormalize $\beta^2 \gamma$.   We found that, indeed, the
prefactor of the relation $\langle r^2(t) \rangle \sim a t^{1-\delta}$
does depend on these irrelevant details.  As we see from
Figure 4, however, the exponent is independent of these details.

The choice of pseudo-random number generator to use is an important
consideration in all Monte Carlo simulations.
A pseudo-random number generator is used in three components of the
present simulation:
in lattice creation, in selecting the
particle initial positions, and in creating the 
transition rates and hop directions.  
The two random number generators employed in our
simulation are a sum of three linear congruential generators \cite{Byte} and an
exclusive-or, linear-feedback-shift register method with table length 55
\cite{Knuth}.
Both of these generators are thought to be fairly reliable.
All of the results in Figures 1-5 were produced by the
sum of three linear congruential generators method.
  Figure 6 compares the mean square displacements produced by these
two generators.
 \begin{figure}[t]
\centering
\leavevmode
\psfig{file=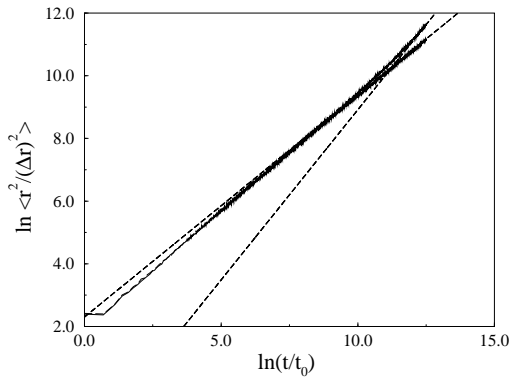,height=2in,angle=-90}
\caption[]
{\label{fig5}
Shown is the mean square displacement versus time for
$\beta^2 \gamma = 10$ and $\hat \chi_{vv}(k) = \gamma  \exp(-k^2/2)/k^2$.
 We have used Eq.
(\ref{5}) to generate the lattice and we placed the random walker
uniformly and randomly at the beginning of each random walk.
The top solid curve comes from using the
exclusive-or, linear-feedback shift-register pseudo-random number
generator \cite {Knuth}, and the bottom solid curve comes from the
 sum of three linear congruential pseudo-random number
generators
\cite{Byte}.  The bottom curve has the expected slope of
 $0.71 = 1-1/(1 + 8\pi / 10)$ at long times.
 The top curve has a slope
of 1.09 at the longest times shown,
 which indicates superdiffusion.  Dashed lines with slopes of 0.71 and 1.09
are shown for convenience.
}
\end{figure}
The three linear
congruential generator method results in
slopes for all values of $\beta^2 \gamma$ that
are consistent with theory.  Interestingly, the linear feedback shift
register method always leads
to slopes greater than expected.  Using this
generator,
 one would conclude that
$\beta^2 \gamma$ is renormalized a finite amount, by some unknown
factor.  In fact, the factor is the inadequacy of the pseudo-random number
generator!  At long times, the exponent of the mean square displacement
exceeds unity. 
This super-diffusive behavior is in conflict with a rigorous bound known
for diffusion in Gaussian random media:
$\lim_{t \rightarrow \infty} \langle r^2(t)/t \rangle \le 4 D \exp[-
\beta^2 \chi_{vv}(0)]$ \cite {deMasi}.  In fact, at long times,
this linear feedback shift register method appears to produce
ballistic behavior, $\langle r^2(t) \rangle \sim a t^2$ (data not
shown).  This
incorrect super-diffusive behavior appears only for potentials that lead to
anomalous diffusion.  The long-range correlations in the potential,
which lead to the anomalous diffusion, apparently couple to the
residual correlations in this pseudo-random number generator.
Similar, poor results from this type of
generator have been observed in another
system with long-range correlations---the Ising model at its critical
point \cite {Ferrenberg}.  In this case, a feedback generator with
a short table length led to predicted critical exponents differing
from the true values by many times the estimated standard deviation.

\section*{6. Conclusion}
We find a satisfying match between theory and numerical results
in our simulation of anomalous diffusion.
  We found that anomalous behavior occurs in the long
time regime, with a transition from normal diffusion at
short times.  We find that the prefactor for the scaling of the
mean square displacement is renormalized by short-distance correlations
in the potential, although the exponent is not.  Reasonable distributions
for the initial conditions lead to the same exponent for the mean 
square displacement.  Interestingly, the correct anomalous diffusion behavior
is observed only with a high quality pseudo-random number generator.
 
\section*{Acknowledgments}
This research was supported by the National Science Foundation
through grants CHE--9705165 and CTS--9702403.

\bibliography{victor}

\end{document}